\newif\iflncs\lncsfalse
\newif\ifmai\maifalse
\newif\ifanon\anonfalse
\iflncs
\documentclass{llncs}
\else
\documentclass[12pt]{article}
\fi
\usepackage{url}
\usepackage[mathscr]{euscript}%
\usepackage[english]{babel}
\usepackage[T1]{fontenc}
\usepackage{amssymb,amsmath}
\usepackage{graphicx}
\usepackage{pgf}
\usepackage{tikz}
\usepackage{wrapfig}
\usepackage{subfig}
\usepackage{geometry}
\usepackage{pspicture}
 \usepackage[usenames,dvipsnames]{pstricks}
\usepackage{epsfig}
 \usepackage{pst-grad} 
 \usepackage{pst-plot} 
\iflncs
\else

\fi
\iflncs\else

\fi

\iflncs{}\else \fi
\iflncs{}\else \fi
\iflncs{}\else \fi

\iflncs{}\else \fi

\iflncs{}\else  \fi

\iflncs{}\else \fi

\iflncs{}\else \fi
\iflncs{}\else \fi

\newcommand{\INDIP}{\hspace{1.5mm}\perp\hspace{-3.9mm}\perp\hspace{1.5mm}}

\newcommand{\defi}{\stackrel{\triangle}{=}}

 \usepackage[utf8x]{inputenc}

\title{Causes of Effects via a Bayesian Model Selection Procedure }

\date{}

\author{ Fabio Corradi$^{\rm a}$ and Monica Musio$^{\rm b}$  \\
\small{$^{\rm
a}$ {\em{Department of Statistics, Computer Science, Applications, University of Firenze, IT }}} \\
\small{$^{\rm b}$ {\em{Department of Mathematics and Computer Science, University of Cagliari, IT}}}}

\begin{document}
\maketitle
\begin{abstract}
In causal inference,  and specifically in the \textit{Causes of Effects} problem, one  is interested in how to use statistical evidence  to understand  causation in an individual case, and so how to assess the so-called {\em probability of causation} (PC).
 The answer relies on the potential responses, which can  incorporate information about what would have happened to the outcome as we had observed a different value of the exposure.
However, even given the best possible statistical evidence for the association between exposure and outcome, we can typically only provide bounds for the PC.
Dawid et al. (2016)  highlighted some fundamental conditions, namely, exogeneity, comparability, and sufficiency, required to obtain such bounds, based on experimental data.
The aim of the present paper is to provide methods to find, in specific cases, the best subsample of the reference dataset to satisfy such requirements.
To this end, we introduce a new variable, expressing the desire to be exposed or not, and we  set the question up as a model selection problem.
The best model will be selected using the marginal probability of the responses and a suitable prior proposal over the model space.
An application in the educational field is presented.
 \end{abstract}


\section{Introduction}

The Causes of Effects (CoE) problem concerns the study of individual causation and  explicitly refers to something happened to a well identified individual.
This nuance of causation has received less attention than the study of the \textit{Effects of  Causes} (EoC), also called the general causation problem.
Actually, in EoC, the aim is the prediction of an outcome  after the realization of an alleged cause; in CoE, instead, we want  to evaluate the \textit{Probability of Causation} (PC), i.e. the probability  the  observed outcome would not have been realized if the alleged cause  had not been made effective, despite the fact that the cause and the outcome were already observed.
Since  we always refer  causation to a specific individual, for the sake of simplicity,  we call her  Ann.

A simplified CoE question is the following.
``Ann had a headache and decided to take aspirin.
Her headache went away.
Was that caused by the aspirin?''

   In CoE, to evaluate the probability of causation,  formally defined in Section \ref{sec:notation}, the relevant questions are:  How might one  use experimental and/or  observational data, gathered on a reference group to which Ann belongs? How would one find  the characteristics shared with Ann by the group of individuals from which  the data came?
These problems, also called  the ``Group to individual'' (G2i) issue in forensic science, have generated a large debate in legal circles (see Faigman {\it et~al}. (2014) \cite{faigman:2014}).
   
   The issue  was extensively studied in the statistical literature  from a technical and a philosophical point of view,  reaching different and somewhat related results.
Essential references are  Dawid (2000; 2016),  \cite{dawid-et-al:2016}, \cite{dawid:2000}  and  Pearl (2009; 2015) \cite{Pearl:2009}, \cite{Pearl:2015}.
   Even if their approaches follow different routes, they  agree about the use of counterfactuals and potential outcomes originating from  Neyman (\cite{neyman:1923}) and re-introduced in the modern literature by Rubin (1974) (\cite{rubin:1974})\footnote{Rubin's approach has been mainly used to solve  EoC-type problems: for this reason it will be not considered in this paper.}.
The need for counterfactuals in CoE is well illustrated by  the question: ``What would be the probability of the response if no treatment  had been provided to the individual in whom we are interested?'' Since only the treatment can be assigned to the individual, the matter concerns a counterfactual, i.e. an event which assumes something different from what happened.
Depending on the assumptions, different results emerge in the forms of a precise probability or bounds  (see Section \ref{sec:review}).
Although not as specific as one would ideally like, such bounds can be of use.
In particular, if the lower bound exceeds one-half, then, in civil cases, we can infer causality ``on the balance of probabilities''.
 To evaluate the PC, we follow the approach of Dawid {\it et~al}.
(2016) \cite{dawid-et-al:2016}, that identified and detailed three {\em fundamental  conditions}  used to estimate from the data  upper and lower bounds for the probability of causation.

A related problem is the choice of the reference population.
More specifically, how to find that group, from among those obtainable from partitioning the randomized experiment sample according to  Ann's characteristics, which best fulfills the fundamental conditions.
 How much information to take into account and how to select the best comparison group  is a tricky issue, also because the choice of the reference class can affect significantly  the conclusions of the inference.
 
The  main aim of the present paper is to make the fundamental conditions  operative by means of empirical testing of such underlying assumptions, in a way that comes up with the ``best'' comparison group.
We set the question up as a Bayesian model selection problem,  where each model specifies a particular choice (more or less detailed) for the characteristics to be included, shared by Ann and the other individuals participating in the study.
The best model  will be selected considering the marginal probabilities of the  response and a satisfactory prior proposal in the model space.
To this end, we introduce a new variable that expresses, for each individual in the study, the desire to receive the treatment or not.
This variable allows introducing the fundamental conditions in the model selection procedure.
Such a method can have various applications in  sociology and education, as well as in medicine.
We present an example in the field of education, where we investigate  the relation between success in a test and whether or not a hint was received (taking into account the student's preference for receiving or not receiving the hint).
 The structure of this paper is as follows.
We first introduce the notation and we define formally the Probability of Causation (PC).
After the review of some  results in the CoE literature, see Section \ref{sec:review}, we provide  the assumptions we require, Section \ref{sec:aim}.
Then, in Section \ref{sec:merge}, we detail how to find the reference sample suitable for evaluating CoE by a model selection procedure.
After presenting the application in Section \ref{sec:application}, we  draw some final conclusions.

\section{Notation}
\label{sec:notation}

We first specify the notation we need  to introduce our approach.
\\
Given data obtained from an ideal large randomized study  concerning $n$ individuals, drawn from a population to which Ann belongs,
 assume that the study  records the outcome  $R=\{0,1\}$ of a treatment $T=\{0,1\}$, which is supposed large enough that the sampling variability of the estimates is negligible.
 Let $H$ be  a  large set of variables $H=\{H_1,\dots,H_k\}$,  characteristic of  both Ann and   the individuals participating in the  study, where we assume  that each $H_j$, $j\in 1,\dots,k$  is discrete (or even dichotomous).
  We denote by $H^{A}$ the value that the set of variables $H$ has  for Ann.
 Since Ann not only  takes the aspirin but also expresses her desire to take it,   we introduce a variable  $E=\{0,1\}$ (for Ann  $E=1$) expressing such a desire for each individual in the study.
This  variable,  first proposed as an unobservable by  Dawid   (2011),  is here considered observable.
 We also introduce the potential variables $R_T=\{0,1\}$,  so that, if the triple $(T,R_0,R_1)$ were observed contemporaneously, it would be easy to solve the CoE problem by stating the probability of causation (PC), defined as
  
  \begin{eqnarray}
\label{eq:PCa}
PC_A& \defi & \Pr(R_0=0|H^A,T=1,R_1=1) \nonumber \\
&=& \dfrac{\Pr(R_0=0,R_1=1|H^A, T=1)}{\Pr(R_1=1|H^A,T=1)}.
\end{eqnarray}
Unfortunately, $R_0,R_1$ are not jointly  estimable from the data, and consequently $PC_A$ cannot be evaluated without some further assumptions.
\section{Some results from the literature}
\label{sec:review}

In epidemiology, $PC_A$ has often been expressed by the quantity  referred to as the \textit{ excess risk ratio} (see for instance \cite{rothman:2012})
\begin{eqnarray}
\label{eq:ERR}
ERR &=&\dfrac{\Pr(R_1=1|H^A)-\Pr(R_0=1|H^A)}{\Pr(R_1=1|H^A)}
\end{eqnarray}
and sometime evaluated in terms of the \textit{observational risk ratio},
\begin{eqnarray}
\label{eq:ORR}
ORR&=& 1- 1/\dfrac{\Pr(R=1|H^A, T=1)}{\Pr(R=1|H^A, T=0)}.
\end{eqnarray}
The quantity (\ref{eq:ORR}), which  plays an important role in the developments we consider, can be evaluated by using the data for the treated and untreated individuals coming from a randomized study or from observational data.
The choice between these two sources  of data depends on the assumptions.
We review the following three  contributions.

\begin{description}
\item[a)]
Pearl, 2000 (Theorem 9.2.14) \cite{Pearl:2000}, showed that, under Exogeneity ($R_0,R_1\INDIP T |H^A$, i.e. the potential outcomes $(R_0, R_1)$ have the same joint distribution, among both treated and untreated study subjects sharing the same background information $H^A$ as Ann) and Monotonicity  ($R_0=1 \rightarrow R_1=1$, i.e.  if Ann were to recover if untreated, she would certainly recover if treated), $PC_A$ is identified and equal to ORR, evaluated on observational data.
This result is remarkable since it ends up with a precise probability.
At the same time, these assumptions 
  are not easily defensible: Exogeneity,  also called \textit{Strong Ignorability}, is  reasonable for data coming from a randomized study but it is  considered weak for observational data.
Monotonicity, also called \textit{No Prevention}, is apparently reasonable (the treatment cannot be obtain worse results than the placebo) but  for some individuals, for example those allergic to a medical treatment, it may not hold.
 In any case, this  latter assumption can not usually be verified.
 \item[b)] Tian and Pearl, 2000 \cite{Tian:2000}  demonstrated  that, relaxing Exogeneity but retaining Monotonicity, the evaluation of  the probability of causation can be obtained in a more refined form as 
 \begin{eqnarray}
PC_A &=&\dfrac{\Pr(R^{obs}=1|T=1,H^A)-\Pr(R^{obs}=0|T=1, H^A)}{\Pr(R^{obs}=1|T=1, H^A)} + \nonumber \\
&+& \dfrac{\Pr(R^{obs}1|T=0, H^A)-\Pr(R^{exp}=1|T=0,H^A)}{\Pr(R^{obs}=1,T=1|H^A)} \label{eq:Tian}
\end{eqnarray}
where $obs$  and $exp$ specify the source of the data (observational or experimental).
 This result points out that  CoE  has both  an experimental and an observational nature.
 Data  from a  randomized experiment amount for no confounding, i.e. the desirable Exogeneity property can be assumed quite safely.
At the same time, Ann made  a choice to receive the treatment, i.e. she was not forced to receive it.
 Hence, a difference between $\Pr(R^{obs}=1| T=0,H^A)$ and $\Pr(R^{exp}=1|T=0,H^A)$ is plausible and must be taken into account.
For instance, if Ann's disease is at an advanced stage, she has  little will to be treated because she perceives that her survival is  almost independent of the treatment.
Since expression (\ref{eq:Tian}) points out the double nature of CoE, its computation requires having data from two different surveys, and this can be problematic.
\item[c)] Dawid  {\it et~al}. (2016) consider the possibility of evaluating CoE by using only data coming from a randomized experiment.
The authors proceed in two steps.
 First they derive bounds for the probability of causation based on the potential outcomes and the constraints implied in their joint distribution.
Interestingly, the relevant $PC_A$ lower bound is equal to ERR (see \eqref{eq:ERR}).
\[ PC_A > \max\{0,1-1/RR_A\}
\]
where $RR_A$, the risk ratio, is 
\begin{equation}
\label{eq:RRa}
RR_A=\frac{\Pr(R_1=1|H^A,T=1)}{\Pr(R_0=1|H^A,T=1)}.
\end{equation}

 Specifically, a large lower bound  is the minimum probability of observing the  response opposite to that actually observed if the treatment were not provided, so \textit{demonstrating} that a different story would have been possible if the treatment had not been applied.
 
  The question is: 
  
``Even accepting working with bounds,  what  cautions must be taken to estimate 
\begin{equation}
\label{eq:comparability}
\Pr(R_1=1|H^A,T=1)
\end{equation}
and
\begin{equation}
\label{eq:sufficiency}
\Pr(R_0=1|H^A,T=1)
\end{equation}
from experimental data?''

 To answer the question, the authors detailed three  conditions, called the {\em fundamental conditions}, to be assumed so as to estimate upper and lower bounds for $PC_A$ based on the marginal probabilities of the response.

The three conditions are the following:

\begin{enumerate}
 \item Exogeneity: Already defined in Section \ref{sec:review}.
 
 \item Comparability:  Ann's potential response, $R^A_1$, is comparable  with those of the treated subjects having the same background characteristics $H^A$ as Ann.
 \item Sufficiency:   Ann's potential response  $R^A_0$ and those of the untreated subjects, all  having the same background characteristics $H^A$ as Ann, are comparable.
\end{enumerate}

\end{description}

While Exogeneity follows directly from randomization, the other two conditions deserve careful reasoning, so the authors restrict their approach to \textit{when we can make good arguments for the acceptability of these fundamental conditions}.

\section{Validating the fundamental conditions}
\label{sec:aim}

Our proposal  to evaluate the CoE  consists in finding,  among all the possible groups of individuals  differing from the specification of $H$,  that one  that ``best fits''  the conditions of Comparability and Sufficiency.
The problem of validating the assumptions is turned into the search for the most suitable group of experimental data supporting the fundamental conditions.

To take into account the experimental and observational nature of the CoE, we consider as \textit{observed}  the variable $E$, (see Section \ref{sec:notation}), 
the decision  to receive the treatment.
For Ann,  this variable  provides  some indirect information about her  state of health,  in the light of which it would no longer be
appropriate to consider her similar to individuals in a pure
experimental study for which this information was not available.
For this reason we extend our experimental data to include the desire of the individuals in the sample to be treated or not.
We believe that it is possible to get this information from people who have accepted a randomized treatment and we also believe that this practice is much less troublesome than to have a double survey of the same population, as in  Tian and Pearl (2000) (\cite{Tian:2000}) (see Section \ref{sec:review}).

In this extended scenario, Comparability means that, conditional on my knowledge of the pre-treatment characteristics of Ann  and the trial subjects,  I regard Ann's potential response as comparable with those of the sub-group identified by ($T=1,E=1$)  having characteristics $H^A$.
In the same framework, the Sufficiency condition refers to the counterfactual scenario in which Ann was not treated.
 In this case we do not have information about Ann's response nor the  information concerning her will to receive the treatment or not.
Apparently, for $T=0$, we could imagine that Ann did not desire receive the treatment (so $E=0$) but it might also be possible that she did not have the drug available, but her wish was to receive it (so $E=1$).
Our concern is to find the specification of $H$ that makes irrelevant the influence of $E$ on the responses in the untreated group.
If we can obtain reasonable support for the condition $R_0 \INDIP E | H^A$,
i.e. if
\begin{equation}
\label{eq:cond2}
\Pr(R_0=1|H^A,  E=1)= \Pr(R_0=1|H^A, E=0), 
\end{equation}
 it would  be possible  to estimate (\ref{eq:sufficiency}) by $\Pr(R=1|H^A, T=0)$ using the data of the untreated.

 \section{Model selection}
 \label{sec:merge}
To perform a selection from  models characterized by different $H$, we need to compute  the marginal probabilities of the observed responses 
of Ann and the individuals participating in the study.
We assume a relevant effect of the treatment that can be modelled using partial exchangeability among treated and untreated individuals.
Furthermore the Comparability condition establishes that we are not able to distinguish between Ann and the group of treated individuals who desire to receive the treatment (identified by $(T=1, E=1)$), sharing with Ann the same characteristics $H^A$, as it concerns the uncertainty of their responses to the treatment.
Since Comparability focusses on the group with the same characteristics as Ann, we don't need to detail the responses of the remaining  individuals in the treated group,  and we model them as exchangeable.

The Sufficiency condition requires that in the untreated group, individuals sharing Ann's characteristics (despite the fact that  for some of them $E=0$ and for others $E=1$) are considered exchangeable.
Also here we are not interested in distinguishing between the remaining individuals in the untreated group (those not sharing Ann's characteristics), so that the untreated group is  modeled as partially exchangeable.
 Of course according to what characteristics are included in  $H$,  different individuals in the randomized sample  will be compared with Ann.
We have $2^k$ ways of selecting a set  of characteristics  from $H$.
Let $J$ be one of these choices,  identified as a subset of $\{1,\ldots,K\}$.
Each choice of $J$ induces a partition of the sample (treated and untreated), which defines a model $M_j$.

Then, by assuming partial exchangeability and by using de Finetti's representation theorem inside every specified exchangeable group, we  can evaluate the probability of observing  Ann and the group of responses, induced by different subsets of $H$.
 
 In this way we turn the issue of finding the group most supporting the fundamental conditions into a model selection problem, solved, as usual,  by computing the marginal  probability of the data conditionally on different instantiations of $H$.

Restricted to the treated group, we denote by $A_{1,1}$ the set of individuals who desire to take the treatment and share the same characteristics as Ann, with $\bar{A}_{1,1}$ its complement, and with  $\mathbf{r}_{A_{1,1}}$, $\mathbf{r}_{\bar{A}_{1,1}}$ the corresponding  vector of responses, while $r_A$ denotes Ann's response.
 In the untreated group, let $A_{0,e}$ be the sets of individuals considered, $e=\{0,1\}$, $A_0=A_{0,0}\cup A_{0,1}$ and  $\bar{A}_{0}$ its complement.
 We extend  these notations in the obvious way to the vectors of responses and to the mixing parameters.
We have:

\begin{align*}
  & \Pr(r_A,\mathbf{r}_{A_{1,1}}, \mathbf{r}_{\bar{A}_{1,1}}, \mathbf{r}_{A_{0,1}}, \mathbf{r}_{A_{0,0}} , \mathbf{r}_{\bar{A}_{0}}) |M_J)= \\
 & \int_{\Theta}\Pr(r_A,\mathbf{r}_{A_{1,1}}, \mathbf{r}_{\bar{A}_{1,1}}, \mathbf{r}_{A_{0,1}}, \mathbf{r}_{A_{0,0}} , \mathbf{r}_{\bar{A}_{0}} \mid \theta_{A_{1,1}},\theta_{\bar{A}_{1,1}}, \theta_{A_{0}},\theta_{\bar{A}_{0}}, M_j)\cdot \\
& \cdot \pi(\theta_{A_{1,1}},\theta_{\bar{A}_{1,1}}, \theta_{A_{0}},\theta_{\bar{A}_{0}})d\theta_{A_{1,1}}d\theta_{\bar{A}_{1,1}}d\theta_{A_{0}}d\theta_{\bar{A}_{0}}= \\
&= \int_{\Theta}\Pr(r_A,\mathbf{r}_{A_{1,1}}, \mathbf{r}_{\bar{A}_{1,1}}, \mathbf{r}_{A_{0,1}}, \mathbf{r}_{A_{0,0}} , \mathbf{r}_{\bar{A}_{0}} \mid \theta_{A_{1,1}},\theta_{\bar{A}_{1,1}}, \theta_{A_{0}},\theta_{\bar{A}_{0}}, M_j)\cdot \\
& \cdot \pi(\theta_{A_{1,1}})\pi(\theta_{\bar{A}_{1,1}})\pi(\theta_{A_{0}})\pi(\theta_{\bar{A}_{0}})d\theta_{A_{1,1}}d\theta_{\bar{A}_{1,1}}d\theta_{A_{0}}d\theta_{\bar{A}_{0}}= \\
& \int_{\Theta_{A_{1,1}}} \Pr(r_A,\mathbf{r}_{A_{1,1}}\mid \theta_{A_{1,1}},M_J)\pi(\theta_{A_{1,1}})d\theta_{A_{1,1}} \cdot  \int_{\Theta_{\bar{A}_{1,1}}} \Pr(\mathbf{r}_{\bar{A}_{1,1}}\mid \theta_{\bar{A}_{1,1}},M_J)\pi(\theta_{\bar{A}_{1,1}})d\theta_{\bar{A}_{1,1}}\cdot \\ \nonumber
&\cdot \int_{\Theta_{A_{0}}} \Pr(\mathbf{r}_{A_{0,1}}, \mathbf{r}_{A_{0,0}}\mid  \theta_{A_{0}},M_J)\pi(\theta_{A_{0}})d\theta_{A_{0}}\cdot \int_{\Theta_{\bar{A}_{0}}} \Pr(\mathbf{r}_{\bar{A}_{0}}\mid  \theta_{\bar{A}_{0}},M_J)\pi(\theta_{\bar{A}_{0}})d\theta_{\bar{A}_{0}}
\end{align*}
By the de Finetti's Representation Theorem, the conditional probabilities of the responses are a mixture of a binomial model and a mixture distribution over the corresponding parameters $\theta$s that are assumed independent of each other.
As a consequence the overall integral easily factorizes.

Now we provide expressions for the above integrals.
All the details are in  Appendix (\ref{app:1}). We have:
\begin{align}
\int_{\Theta_{A_{1,1}}} \Pr(r_A,\mathbf{r}_{A_{1,1}}\mid \theta_{A_{1,1}},M_J)\pi(\theta_{A_{1,1}})d\theta_{A_{1,1}}=\frac{x_{A_{1,1}}+1}{n_{A_{1,1}}+2} \cdot \frac{1}{n_{A_{1,1}}+1}
\label{eq:margsuff1}
\end{align}

\noindent where  $x_{A_{1,1}}$ is the number of successes in the group $A_{1,1}$ and $n_{A_{1,1}}=|A_{1,1}|$.
The notation is extended in the obvious way in the other groups.

\begin{align}
\int_{\Theta_{\bar{A}_{1,1}}} \Pr(\mathbf{r}_{\bar{A}_{1,1}}\mid \theta_{\bar{A}_{1,1}},M_J)\pi(\theta_{\bar{A}_{1,1}})d\theta_{\bar{A}_{1,1}}=\frac{1}{n_{\bar{A}_{1,1}}+1}
\label{eq:margsuff2}
\end{align}

\begin{align}
\int_{\Theta_{A_{0}}} \Pr(\mathbf{r}_{A_{0,1}}, \mathbf{r}_{A_{0,0}}\mid  \theta_{A_{0}},M_J)\pi(\theta_{A_{0}})d\theta_{A_{0}} &=\dbinom{n_{A_{0,0}}}{x_{A_{0,0}}}\dbinom{n_{A_{0,1}}}{x_{A_{0,1}}}\dbinom{n_{A_{0,0}}+n_{A_{0,1}}}{x_{A_{0,0}}+x_{A_{0,1}}}^{-1}\cdot \\
& \cdot\frac{1}{n_{A_{0,0}}+n_{A_{0,1}}+1} 
\label{eq:margsuff3}
\end{align}

\begin{align}
\int_{\Theta_{\bar{A}_{0}}} \Pr(\mathbf{r}_{\bar{A}_{0}}\mid
 \theta_{\bar{A}_{0}},M_J)\pi(\theta_{\bar{A}_{0}})d\theta_{\bar{A}_{0}}= \frac{1}{n_{\bar{A}_{0}}+1}
\label{eq:margsuff4}
\end{align}

\begin{figure}[h] \centering
\includegraphics[scale=0.8]{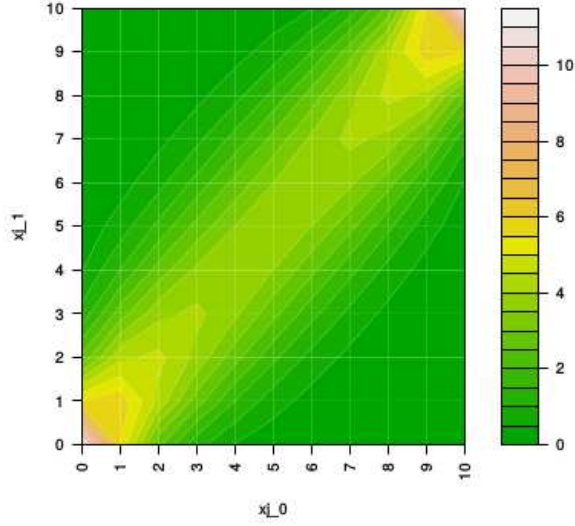} 
\caption{Hypergeometric behaviour for $n_{A_{0,0}}=n_{A_{0,1}}=10$ and different values for $x_{A_{0,0}}$ and $x_{A_{0,1}}$ }
\label{figure:1}
\end{figure}

Readers might recognize the conditional  Irving--Fisher exact test as part  of  (\ref{eq:margsuff3}).
The result is not surprising since we are looking for the set of $H$ making $E$ irrelevant, and thus supporting the hypothesis of no-difference between the success ratio in the two groups.
An illustration of the behavior of the hypergeometric for $n_{A_{0,0}}=n_{A_{0,1}}=10$ is given in Figure \ref{figure:1}.
High support to the model is achieved when the number of successes $x_{A_{0,0}}$ and $x_{A_{0,1}}$ is almost the same in the two groups.

\subsection{Prior and posterior in the model space}
\label{sec:comb}
 The goal is to evaluate the posterior probability  of $M_J$ given the responses observed for Ann and the sample.
The overall marginal likelihood of $M_j$ is the product of (\ref{eq:margsuff1}), (\ref{eq:margsuff2}), (\ref{eq:margsuff3}) and (\ref{eq:margsuff4}) 
\small{
\begin{align}
\label{eq:margianlik3}
& \Pr(r_A,\mathbf{r}_{A_{1,1}}, \mathbf{r}_{\bar{A}_{1,1}}, \mathbf{r}_{A_{0,1}}, \mathbf{r}_{A_{0,0}} , \mathbf{r}_{\bar{A}_{0}}) |M_J)= 
    \frac{x_{A_{1,1}}+1}{n_{A_{1,1}}+2} \cdot \frac{1}{n_{A_{1,1}}+1}\cdot\frac{1}{n_{\bar{A}_{1,1}}+1}\\ \nonumber
   \cdot & \dbinom{n_{A_{0,0}}}{x_{A_{0,0}}}\dbinom{n_{A_{0,1}}}{x_{A_{0,1}}}\dbinom{n_{A_{0,0}}+n_{A_{0,1}}}{x_{A_{0,0}}+x_{A_{0,1}}}^{-1}\frac{1}{n_{A_{0,0}}+n_{A_{0,1}}+1} \cdot \frac{1}{n_{\bar{A}_{0}}+1}
  \end{align}
  \normalsize{
Concerning the prior over the space of models, the simplest choice is to consider a uniform distribution }
\begin{equation}
   \label{eq:uniprior} \Pr(M_J)=\frac{1}{2^k}.
   \end{equation}
Another choice is the one proposed by 
Chen and Chen, (2008) \cite{chen:2008}.
They give the same prior probability (equal to $\frac{1}{k+1}$) for all models sharing the same number of characteristics $k$.
In this way, for the generic model $M_J$, we have
\begin{equation}
   \label{eq:chenprior}
   \Pr(M_J)= \frac{1}{k+1}\binom{k}{|M_J|}^{-1}\cdot I(|M_J|\leq k/2)
 \end{equation}
where  the search spans all models including at most $k/2$
 characteristics.
This last  choice favors model selection according to 
 Occam's razor principle: the fewer characteristics employed, the more probable is the model.
 This rationale is reasonably objective.
Combining (\ref{eq:uniprior}) or (\ref{eq:chenprior})
with  (\ref{eq:margianlik3}), we get the required posterior.

\subsubsection*{Computational issues}
If the model size becomes huge, we may not be in a
position to evaluate the normalizing constant of the posterior distribution,
but we can establish an MCMC to make an inference about the variable $M_J$.

Essentially a Metropolis--Hastings would suffice, the acceptance ratio
being given by (\ref{eq:margianlik3}),  evaluated for two different
elements of $M_J$, taking into account the probability of proposing a new model.

\section{Application}
\label{sec:application}
 We  carried out an experiment at the University of Florence, School of Engineering, Fall 2017.
We asked 161 students to solve a simple probabilistic question and we provided randomly a hint (the treatment $T$).
Table (\ref{tab:tab01}) presents the students' background information included in the analysis (the characteristics $H$).
Before the test, we asked the  students if they wished to be helped or not (the desire variable $E$).\\
 
\begin{table}
\caption{List of student characteristics included in the experiment\label{tab:tab01} }
\centering
\fbox{%
 \begin{tabular}{|  l| l  | l |}
 \hline
 \em & \textit{Variable} &  \textit{Description} \\
1 & Engineer Course & 0=Civil, 1= Facilities\\
2 & Gender  & 0=male, 1=female\\
3 & Age  & 0=$>21$, 1=$\geq 21$ \\
4 & Place of birth   & 0= outside Florence, 1= Florence \\
5& Place of residence & 0= outside Florence, 1= Florence \\
6 & Year of Diploma & 0=before 2016, 1=2016 \\
7 & Place of Diploma  & 0= outside Florence, 1= Florence\\
8 & High school   & 0= Other, 1= Technical \\
9 & Diploma vote & 0= <80, 1= $\geq 80$ \\
10 & $1^{st}$ registration at University & 0= before 2016, 1=2016 \\ 
11 & Statistical background   & 0= no, 1=yes \\
12 & Father's level of education  &  $0=\leq$ high school, 1=University\\
13 & Mother's level of education  &  $0=\leq$ high school, 1=university\\
14 & Working student   & 0=no, 1=yes \\
 \hline
\end{tabular}}
\end{table}
 
We wish to investigate whether there is a causal relation between the hint and the ability to solve the question, for those students who desired to receive a hint.
We had 8  of these cases.
The corresponding risk ratio (obtained considering the model best fitting the fundamental conditions)  lies in the interval $[1.73, 2.43]$ and in one case it exceeds $2$ ($RR=2.43$), which
 is a clue for there being a causal relation.
In all the other cases, the causal relation is not strongly supported, since $RR$ is close but does not reach $2$ (see, in Figure (\ref{fig:daghor})), the left side of each sub-picture).
In the right side of each sub-picture in Figure(\ref{fig:daghor}) we illustrate how the model selection procedure proposes models respecting the fundamental conditions.
For the model with the highest probability, the red dot indicates, in the $x-$axis,  the success ratio for treated and, in the $y-$axis, the ratio between the success ratio for untreated with $E=1$ and  with $E=0$, respectively.
These are the main forces driving the marginal probability for the responses, as shown in \eqref{eq:margianlik3}.
Ideally, comparability and sufficiency are mostly supported by the highest possible success ratio among the treated and by a ratio between untreated with E=0 and E=1 approximately equal 1.
 As is apparent, the selected model  achieves a good compromise between these requirements.
The set of variables selected in the 8 cases are shown in Table (\ref{tab:02}).
Note that, overall, these models include only 5 characteristics, and all of them include the educational status of the family.
In 4 out of the 8 cases, the same model, including family education, $1^\text{first}$ University registration (a proxy for  understanding whether the student failed earlier in their educational career)  and previous exposure to statistical training was selected.

\begin{figure}[h]
\includegraphics[scale=0.8]{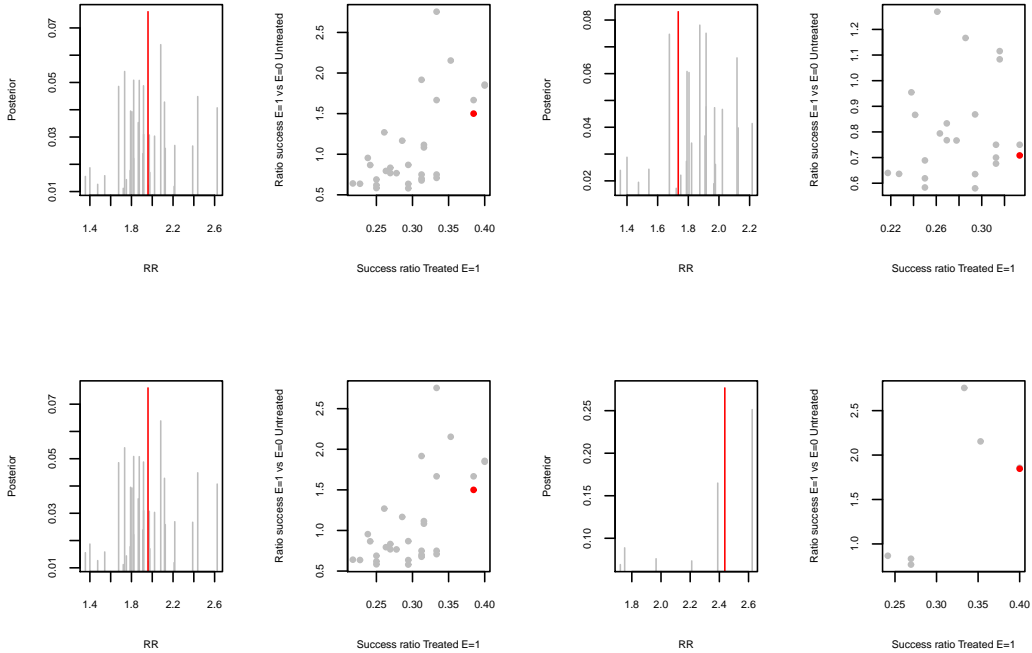} 
\caption{Educational causation.
Risk Ratio and posterior probability for each explored model (left side).
Success ratio for treated vs relative success ratio for untreated with $E=1$ and $E=0$ (right side).
In the picture there is an example concerning four students  who  succeed.
}
\label{fig:daghor}
\end{figure}

\begin{figure}[t] \centering 
\includegraphics[scale=0.8]{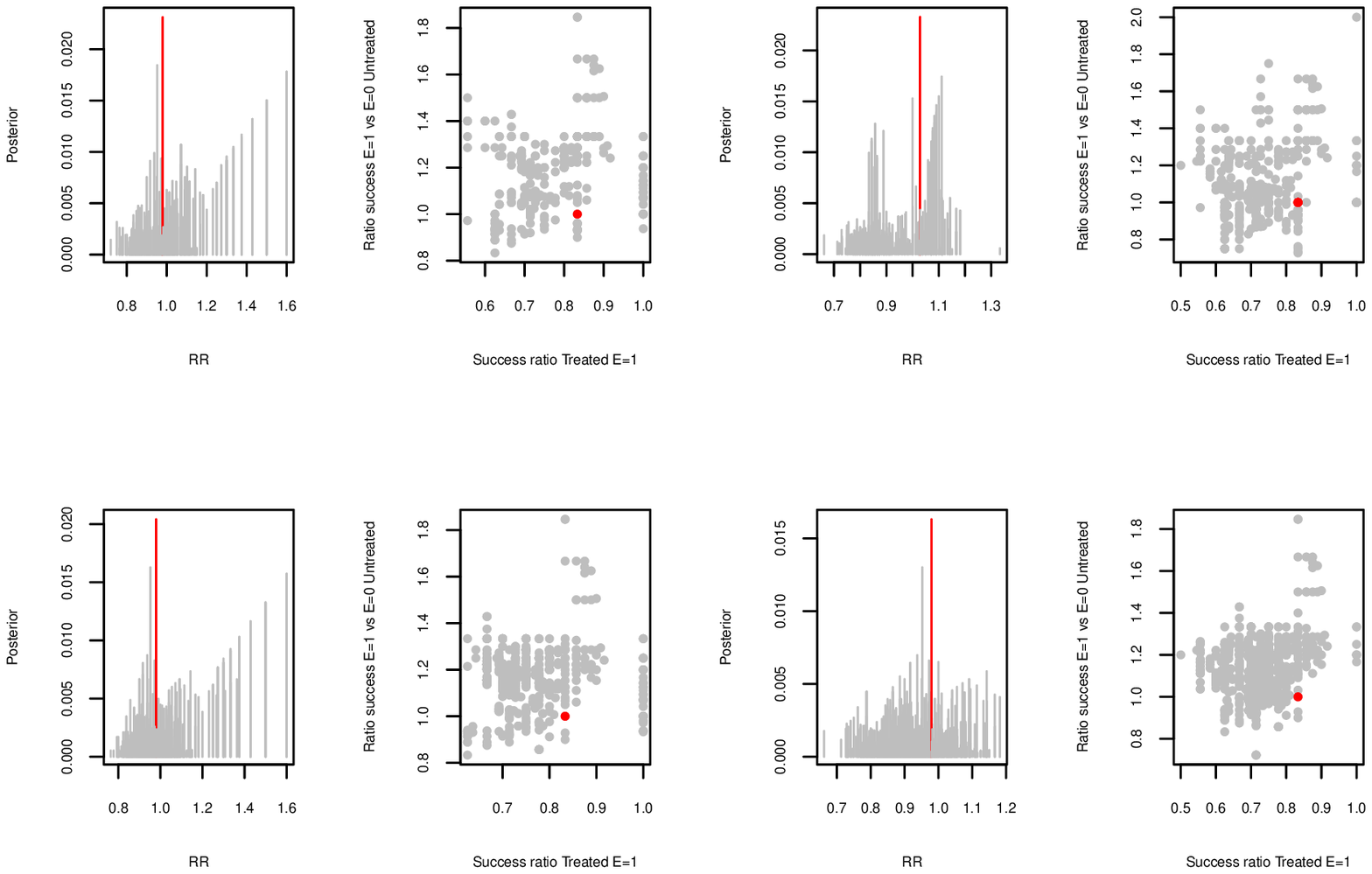} 
\caption{Forensic causation.
Risk Ratio and posterior probability for each explored model (left side).
Success ratio for treated vs relative success ratio for untreated with $E=1$ and $E=0$ (right side).
In the picture there is an example concerning four students  who DID NOT succeed.}
\label{fig:daghor1}
\end{figure}

\begin{table}
\caption{\label{tab:02} Educational causation. For the 8 students who desired the hint, got it, and succeed, there are reported the characteristics $H$ selected by the model that best supported the fundamental conditions and the Risk Ratio}
\centering
\fbox{%
\begin{tabular}{*{10}{c}}
Student & $1^{st}$ University & Statistical & \multicolumn{2}{c}{Education}  & Work & $RR$ \\
 &  registration \em &  background &  Father &  Mother &   &  \\
 \hline
1 & before $2016$ &  no & university & university &  $-$ & $1.96$ \\
2 & before $2016$  & $-$ & university & university &  $-$ & $1.73$ \\
3 & $2016$ & no & university & university &  $-$ &  $1.96$ \\
4 &  $-$ &  $-$ & university &university & yes & $2.43$ \\
5 &before $2016$ &  $-$ & university &university & yes  & $1.96$ \\
6 & $2016$  & no & university &university &   $-$ &  $1.96$  \\
7 &  $2016$  &  $-$ & university &university &  $-$&  $1.96$  \\
8 &  $2016$  & no & university &university &  $-$ &  $1.96$   \\
\hline
\end{tabular}}
\end{table}

As a result of our experiment we also have that among the students who asked for and received the hint, 24 did not succeed.
We can suppose that some of them claimed that it was the hint which caused their failure.
 In this case, the $RR$ lies in the interval $[0.86,1.60]$, an example for four students is in Figure (\ref{fig:daghor1}).
It is not conclusive that there is a causal relation between the  hint and the failure  since all the models for all the considered students provided values of RR much smaller than 2.
 In a civil trial this would not suggest to a judge that compensation be awarded.

\newpage

\section{Conclusions}
\label{sec:conclusions}

 We introduced a typical Causes of Effects problem by means of an archetypical example considering Ann and the effect of an aspirin on her headache.
We have proposed a possible solution to make operational the choice of
variables to include, so as to validate the fundamental assumptions
underlying the assessment of Ann's probability of causation.
 We assume it is possible to take a randomized sample from  Ann's population where, as usual, $T$ is assigned following a randomized protocol and $E$ (this is a novelty) is  a question asking the members of the sample  about their preference to be treated or not.

In the evaluation of $RR_A$ (see \ref{eq:RRa}), an extreme position would be to include all the subjects participating in the experiment so that simply belonging to the reference population
would make the individuals in the sample similar to Ann.
On the other hand,
the choice could be to find the persons most similar to Ann,
i.e. those matching all the available  characteristics.
Clearly
neither of these positions is safe: the former ignores some
characteristics of Ann which could be very influential on her reaction
to the headache after taking aspirin.
 The latter greatly reduces the
number of individuals to be employed in the estimation, so producing
a very unstable inference.
 Our approach takes a sensible middle course and provide a sensible different causal inference for  individuals experiencing the same treatment.
Interestingly, this is exactly the aim of Precision Medicine (see Mesko (2017), \cite{mesko:2017}) which looks for different medical interventions for a group of individuals sharing some relevant (for the reaction between treatment and outcome) characteristics.

The next step will be to extend the method to observational studies, to make possible in a wider range of cases the evaluation of the $PC_A$ for Causes of Effects problems.

\newpage
\newpage 
\section{Appendix}
\label{app:1}
Now we detail the computation of the integrals (\ref{eq:margsuff1}), (\ref{eq:margsuff2}), (\ref{eq:margsuff3}) and (\ref{eq:margsuff4}). We always assume a no informative prior for all $\pi(\cdot)=Beta(1,1)$.
 We start with  (\ref{eq:margsuff1}).
\begin{align}
& \int_{\Theta_{A_{1,1}}} \Pr(r_A,\mathbf{r}_{A_{1,1}}\mid \theta_{A_{1,1}},M_J)\pi(\theta_{A_{1,1}})d\theta_{A_{1,1}}= \nonumber \\
&= \int_{\Theta_{A_{1,1}}}  \Pr(r_A | \mathbf{r}_{A_{1,1}},\theta_{A_{1,1}},M_J)\underbrace{\Pr(\mathbf{r}_{A_{1,1}}|\theta_{A_{1,1}},M_J)\pi(\theta_{A_{1,1}}|M_J)}_{\text{numerator of the }\theta_{A_{1,1}} \text{ updating}}d \theta_{A_{1,1}} \nonumber \\
&= \underbrace{\int_{\Theta_{A_{1,1}}} \Pr(r_A|\theta_{A_{1,1}},M_J)\pi(\theta_{A_{1,1}}|\mathbf{r}_{A_{1,1}},M_J)d\theta_{A_{1,1}}}_{\dfrac{x_{A_{1,1}}+1}{n_{A_{1,1}}+2}}\cdot  \nonumber \\
& \cdot  \underbrace{\int_{\Theta_{A_{1,1}}}\Pr(\mathbf{r}_{A_{1,1}}|\theta_{A_{1,1}},M_J)\Pr(\theta_{A_{1,1}}|M_J)d\theta_{A_{1,1}} }_{\dbinom{n_{A_{1,1}}}{x_{A_{1,1}}}\dfrac{\Gamma(1+1)}{\Gamma(1)\Gamma(1)}  \dfrac{\Gamma(1+x_{A_{1,1}}) \Gamma(n_{A_{1,1}}+ 1 -x_{A_{1,1}})}{\Gamma(n_{A_{1,1}}+ 1 +1)}} \nonumber \\ & =\frac{x_{A_{1,1}}+1}{n_{A_{1,1}}+2} \cdot \frac{1}{n_{A_{1,1}}+1}. \nonumber
\end{align}

\noindent For (\ref{eq:margsuff2})  we have

\begin{align*}
&\int_{\Theta_{\bar{A}_{1,1}}} \Pr(\mathbf{r}_{\bar{A}_{1,1}}\mid \theta_{\bar{A}_{1,1}},M_J)\pi(\theta_{\bar{A}_{1,1}})d\theta_{\bar{A}_{1,1}}= \\ 
&  \underbrace{\int_{\Theta_{\bar{A}_{1,1}}}\Pr(\mathbf{r}_{\bar{A}_{1,1}}|\theta_{\bar{A}_{1,1}},M_J)\Pr(\theta_{\bar{A}_{1,1}}|M_J)d\theta_{\bar{A}_{1,1}} }_{\dbinom{n_{\bar{A}_{1,1}}}{x_{\bar{A}_{1,1}}}\dfrac{\Gamma(1+1)}{\Gamma(1)\Gamma(1)}  \dfrac{\Gamma(1+x_{\bar{A}_{1,1}}) \Gamma(n_{\bar{A}_{1,1}}+ 1 -x_{\bar{A}_{1,1}})}{\Gamma(n_{\bar{A}_{1,1}}+ 1 +1)}} \nonumber \\
& = 
\frac{1}{n_{\bar{A}_{1,1}}+1}
\end{align*}

We now compute (\ref{eq:margsuff3})  

\begin{align*}
&\int_{\Theta_{A_{0}}} \Pr(\mathbf{r}_{A_{0,1}}, \mathbf{r}_{A_{0,0}}\mid  \theta_{A_{0}},M_J)\pi(\theta_{A_{0}},M_J)d\theta_{A_{0}}= \\
&= \int_{\Theta_{A_{0}}} \Pr(\mathbf{r}_{A_{0,0}},|\theta_{A_{0}},\mathbf{r}_{A_{0,1}},M_J)\underbrace{\Pr(\mathbf{r}_{A_{0,1}}|\theta_{A_{0}}, M_J)\pi(\theta_{A_{0}}|M_J)}_{\text{numerator of the }\theta_{A_{0}} \text{ updating}}d\theta_{A_{0}} \nonumber \\
&= \int_{\Theta_{A_{0}}} \Pr(\mathbf{r}_{A_{0,0}}|\theta_{A_{0}},M_J)\pi(\theta_{A_{0}}| \mathbf{r}_{A_{0,1}},M_J)d\theta_{A_{0}} \int_{\Theta_{A_{0}}}\Pr(\mathbf{r}_{A_{0,1}}|\theta_{A_{0}},M_J)\pi(\theta_{A_{0}}| M_J)d\theta_{A_{0}} \nonumber\\
&= \dbinom{n_{A_{0,0}}}{x_{A_{0,0}}} \dfrac{\Gamma(n_{A_{0,1}}+2)}{\Gamma(x_{A_{0,1}}+1)\Gamma(n_{A_{0,1}}- x_{A_{0,1}}+1)}  \dfrac{\Gamma(x_{A_{0,1} } +1 + x_{A_{0,0}})\Gamma(n_{A_{0,0}}+ n_{A_{0,1}}- x_{A_{0,1} }+1- x_{A_{0,0}}) }{\Gamma(n_{A_{0,0}} + n_{A_{0,1}}+2)} \nonumber\\
&\cdot \dbinom{n_{A_{0,1}}}{x_{A_{0,1}}}\dfrac{\Gamma(1+1)}{\Gamma(1)\Gamma(1)}  \dfrac{\Gamma(1+x_{A_{0,1}}) \Gamma(n_{A_{0,1}}+ 1 -x_{A_{0,1} })}{\Gamma(n_{A_{0,1}} + 1 +1)} \nonumber \\
& = \dbinom{n_{A_{0,0}}}{x_{A_{0,0}}}\dbinom{n_{A_{0,1}}}{x_{A_{0,1}}}\dbinom{n_{A_{0,0}}+n_{A_{0,1}}}{x_{A_{0,0}}+x_{A_{0,1}}}^{-1}\frac{1}{n_{A_{0,0}}+n_{A_{0,1}}+1.} 
\end{align*}

Concerning the last term,  (\ref{eq:margsuff4}), with a similar computation as for  (\ref{eq:margsuff2}),   it is straightforward to see that 

\begin{align}
\int_{\Theta_{\bar{A}_{0}}} \Pr(\mathbf{r}_{\bar{A}_{0}}\mid
 \theta_{\bar{A}_{0}},M_J)\pi(\theta_{\bar{A}_{0}})d\theta_{\bar{A}_{0}}= \frac{1}{n_{\bar{A}_{0}}+1}.
\end{align}\\
{\bf Acknowledgement:}{\em The second author was supported by the project GESTA of the Fondazione di Sardegna and Regione Autonoma di Sardegna}

\bibliography{recrutement}
\bibliographystyle{chicago}

\clearpage

\end{document}